\documentclass{aa}  
\usepackage{graphicx}
\usepackage{txfonts}
\usepackage{natbib,twoopt}
\bibpunct{(}{)}{;}{a}{}{,}             
\usepackage{hyperref}
\hypersetup{
    colorlinks=true,
    linkcolor=blue,
    citecolor = blue,
    filecolor=magenta,
    urlcolor=blue}
\usepackage{booktabs}

\begin{document}

\authorrunning{Ferraro et al.}
\titlerunning{GW sources from BFF}

\title{Bulge Fossil Fragments as a new population of factories of gravitational wave sources in the Galaxy}

\author{F. R. Ferraro\inst{1,2}, E. Vesperini\inst{3}, B. Lanzoni\inst{1,2}, D. Romano\inst{2}, L. Origlia\inst{2}, C. Pallanca\inst{1,2}, C. Fanelli\inst{2}, F. Calura\inst{2}, E. Dalessandro\inst{2}, D. Massari\inst{2}, G. Zullo\inst{1,2}, M. Cadelano\inst{1,2}    
}

\institute{ Dipartimento di Fisica e Astronomia, Università degli Studi di Bologna, Via Gobetti 93/2, I-40129 Bologna, Italy \email{francesco.ferraro3@unibo.it} \and
   INAF, Osservatorio di Astrofisica e Scienza dello Spazio di Bologna, Via Gobetti 93/3, I-40129 Bologna, Italy
   \and 
Dept. of Astronomy, Indiana University, Bloomington, IN 47401, USA\\}

\abstract{
The discovery of the complex stellar populations hosted in two massive stellar systems in the Galactic bulge, namely Terzan 5 and Liller 1, posed intriguing questions about their origin and their possible connection with the formation and early evolution of the bulge itself.  Indeed, despite their globular cluster appearance, they host sub-populations with significantly different ages (several Gyrs) and metallicities (about 1 dex) tracing a chemical abundance pattern that is consistent only with that observed in the bulge. These surprising properties can be naturally explained in the context of a self-enrichment scenario, opening the fascinating possibility that they could be the remnants of primordial massive structures that contributed to the bulge formation (the so-called Bulge Fossil Fragments, BFFs) capable of retaining supernova ejecta within their potential well. 
In this paper we present a first attempt to quantify the expected contribution of BFFs to the gravitational wave emission. In particular, by adopting Terzan 5 as prototype of BFF, using its chemical evolutionary model, and following a scaling relation derived for globular clusters, we present a first-guess estimate of the number of binary black hole mergers expected in
this stellar system. Within the adopted simplifying assumptions and the uncertainties about the initial conditions of the proto-Terzan 5 system, we find that several hundreds of binary black hole mergers are expected, a number that is  between $\sim 15$ and $\sim 250$ times larger than that produced by a typical globular cluster.

Hence, this study identifies in the BFF family a new population of stellar systems potentially able to produce a significant number of gravitational wave  emitters, that has not been considered in any previous investigation of gravitational wave sources. Moreover, considering the deep potential well and the high collisional rate of these systems, we speculate that they could also be the natural place where black holes with masses above 60 M$_\odot$ and even intermediate-mass black holes can form via repeated dynamical interactions. 
}


\maketitle

\section{Introduction}
The ongoing photometric, spectroscopic and kinematic characterization of Galactic stellar systems that we started a couple of decades ago (see, e.g., \citealt{origlia02,origlia03,valenti10, lanzoni07,lanzoni18,
leanza22,leanza23,leanza24,libralato18,dalessandro24,ferraro09,ferraro18a,ferraro18b,ferraro25,saracino15,saracino16,saracino19, cadelano22, cadelano23, pallanca19, pallanca21, pallanca23, deras23,deras24,loriga25}) has revealed the existence of a few intriguing and unexpected objects in the bulge. 

In fact, Terzan 5 and Liller 1, which have been considered as genuine globular clusters (GCs) for more than 50 years, have been found to show the signatures of a much more complex formation and evolutionary history. 
The analysis of proper motion-selected and differential reddening-corrected color-magnitude diagrams has shown that, in both systems, an old stellar population of 12-13 Gyr cohabits with a more centrally concentrated and much younger component (of ~4.5 Gyr in Terzan 5 and 1-2 Gyr in Liller 1; \citealt{ferraro16, ferraro21}).
Detailed spectroscopic investigations of the stellar content in these systems \citep{origlia11,massari14,crociati23,fanelli24} indicate that the main and oldest component has iron abundance [Fe/H] $= -0.3$, while the youngest sub-population has super-solar metallicity ([Fe/H] $\simeq +0.3$; note that not a single genuine Galactic GC or local dwarf galaxy shows such a high metallicity). A further (minor) component with [Fe/H] $= -0.79$ has been discovered in Terzan 5  \citep{origlia13,origlia19}. The [$\alpha$/Fe]-[Fe/H] pattern drawn by the sub-populations with measured O, Mg and Si abundance in Terzan 5 \citep{origlia11, origlia13, origlia19, origlia25} is impressively similar to that observed in Liller 1 \citep{crociati23,alvarez24,fanelli24,ferraro25} and clearly shows that the oldest stellar populations (at sub-solar metallicity) formed from gas enriched exclusively by  Type II supernovae (SNeII), while the youngest components have been enriched also by Type Ia supernovae (SNeIa; see, e.g., \citealp{romano23}). Intriguingly, this pattern is fully consistent with that observed for bulge field stars \citep{johnson14} and clearly incompatible with those of the Galactic disk and halo, and any dwarf galaxy in the Local Universe \citep{tolstoy09}. Indeed, as discussed in \citet{origlia25} and \citet{ferraro25}, this pattern unambiguously indicates a strong parental link between the two systems and the bulge, even in the presence of some specific elemental abundance differences \citep{taylor22} due to the fact that Terzan 5, Liller 1 and the bulge followed different specific evolutionary paths. The kinship with the bulge is further testified by the observed kinematics: the reconstructed orbits of both systems \citep{vasil2021} show that they spent their entire life in the bulge, thus  excluding that they are the result of a recent accretion event. The preliminary star formation histories reconstructed for these systems \citep{dalessandro22, crociati24} show a main burst occurred 12-13 Gyr ago (which could be quite prolonged in time), followed by a continuous star formation with very low intensity and with the most recent burst that occurred 1-2 Gyr ago in Liller 1, and $\sim 4$ Gyr ago in Terzan 5. This kind of star formation history is totally incompatible with that of genuine GCs \citep[e.g.,][]{krause+16, calura+19, wirth+24}.

The super-solar metallicity of the young component observed in both systems is larger than that measured in any Galactic GC, thus ruling out the possibility \citep{pfeffer21} that they formed from the merger of two such systems. This scenario, as well as the accretion of a giant molecular cloud by a genuine GC \citep{mckenzie18, bastian22}, is also inconsistent with the observed star formation history. In fact, these are expected to be rare events, which can occur at most once in the lifetime of a stellar system, while Terzan 5 and Liller 1 seem to have experienced a  prolonged or multi-burst star formation activity. The hypothesis that they are the former nuclear star clusters of an accreted dwarf galaxy \citep{brown18, alfaro19} is at odds with their bulge-confined orbit and with the observed [$\alpha$/Fe]-[Fe/H] pattern, which is incompatible with that of dwarf galaxies and in agreement only with that of the Galactic bulge.   
Thus, all the observational results collected so far show that Terzan 5 and Liller 1 are complex stellar systems disguised under the false appearance of GCs, and clearly point to the intriguing possibility that they experienced self-enrichment processes and are deeply linked to the Galactic bulge. In fact, the age (12-13 Gyr) and the chemical composition ([Fe/H] $=-0.3$ and [$\alpha$/Fe]$=+0.35$) of the oldest sub-population turns out to be remarkably similar in Terzan 5 and Liller 1 \citep{ferraro21}, and indicate that these systems formed at the same epoch of the Milky Way formation, from the same (highly metal-enriched and $\alpha$-enhanced) material from which the proto-bulge assembled. The fact that the young sub-populations are more metal-rich and more centrally segregated than the old components \citep{lanzoni10, ferraro21} is suggestive of a self-enrichment scenario and indicates that the progenitors of these systems were massive enough to retain the (iron enriched) SN ejecta within their potential wells. 
Thus, Terzan 5 and Liller 1 could be the relics of two primordial clumps (similar to those observed in clumpy galaxies at $z>2$; see, e.g., \citealt{elme09a,elme09b}) that, at the epoch of the Milky Way assembly, contributed to the formation of the Galactic bulge \citep[e.g.,][]{imm04, elme08, bournaud09}, the first discovered ``Bulge Fossil Fragments''  (BFFs, as named by \citealt{ferraro21}). Of course, most of these primordial structures merged to form 
the Milky Way spheroid, but the densest portion of a few of them could have survived the total destruction \citep{bournaud16} and be now observable under the false identity of GCs.

In the framework of a self-enrichment scenario, the BFFs should be the most efficient factories of compact objects in the Galaxy. In fact, a large number of neutron stars (NSs) and black holes (BHs) is expected to be formed from the large number of SNII explosions needed to explain the chemical patterns observed in the systems. The compact objects retained within the potential well are likely to become part of binary systems that arehardened by stellar interactions. Indeed, Terzan 5 and Liller 1 have the largest collision rate among all Galactic GCs \citep{verbunt87,lanzoni10}. This could also naturally explain the huge reservoir of X-ray sources and millisecond pulsars detected in Terzan 5 \citep[e.g.,][]{ransom05, heinke06, cadelano18, pad24}\footnote{Indeed, $\sim 15\%$ of the entire population of millisecond pulsars observed in Milky Way GCs is found in Terzan 5 \citep{ransom05, cadelano18, pad24}.}

Binary systems hosting degenerate objects as BHs and NSs are the most powerful gravitational wave (GW) sources and, so far, the only ones detectable with the current generation of instruments. Since the first detection \citep{abbott+16}, several events involving BH-BH systems have been observed  \citep{abbott+23a} while only a few of them involved NS-NS systems (such as GW170817, \citealt{abbott+17}, and GW190425, \citealt{abbott+20}) and BH-NS systems \citep[][see also \citealp{abbott+23b}]{abbott+21}. 

Several recent papers (e.g., Choksi et al. 2019; Kremer et al. 2019; Samsing et al. 2019, and reference therein) indicate GCs as the most promising factories of GW sources through binary BH (BBH) mergers, because of their intense internal dynamical activity. Taking into account the high production of binary systems with degenerate companion that is predicted for BFFs, the aim of the present paper is to provide a first estimate of the contribution of this new class of stellar systems to GW emission. To this purpose we focus on Terzan 5, taking advantage of the chemical evolutionary model specifically developed for this system \citep{romano23} and using the relation of \citet{hong18} quantifying the expected number of BBH mergers in GC-like systems.

\section{BH and NS remnants in Terzan 5: constraints from its chemical evolution}
\citet{romano23} have recently presented a chemical evolutionary model demonstrating that the self-enrichment of a stellar system with a total  initial mass of a few $10^7 {\rm M}_{\odot}$ can reproduce all the chemical abundances and patterns observed in Terzan 5. 
They used numerical models \citep{romano13, romano15} that solve the classical set of equations of chemical evolution \citep{tinsley80, matteucci21} including stellar yields for stars with different initial masses and chemical compositions \citep{karakas10, doherty14a, doherty14b, nomoto13, iwamoto99}. 

\begin{figure}
    \centering \includegraphics[width=\columnwidth]{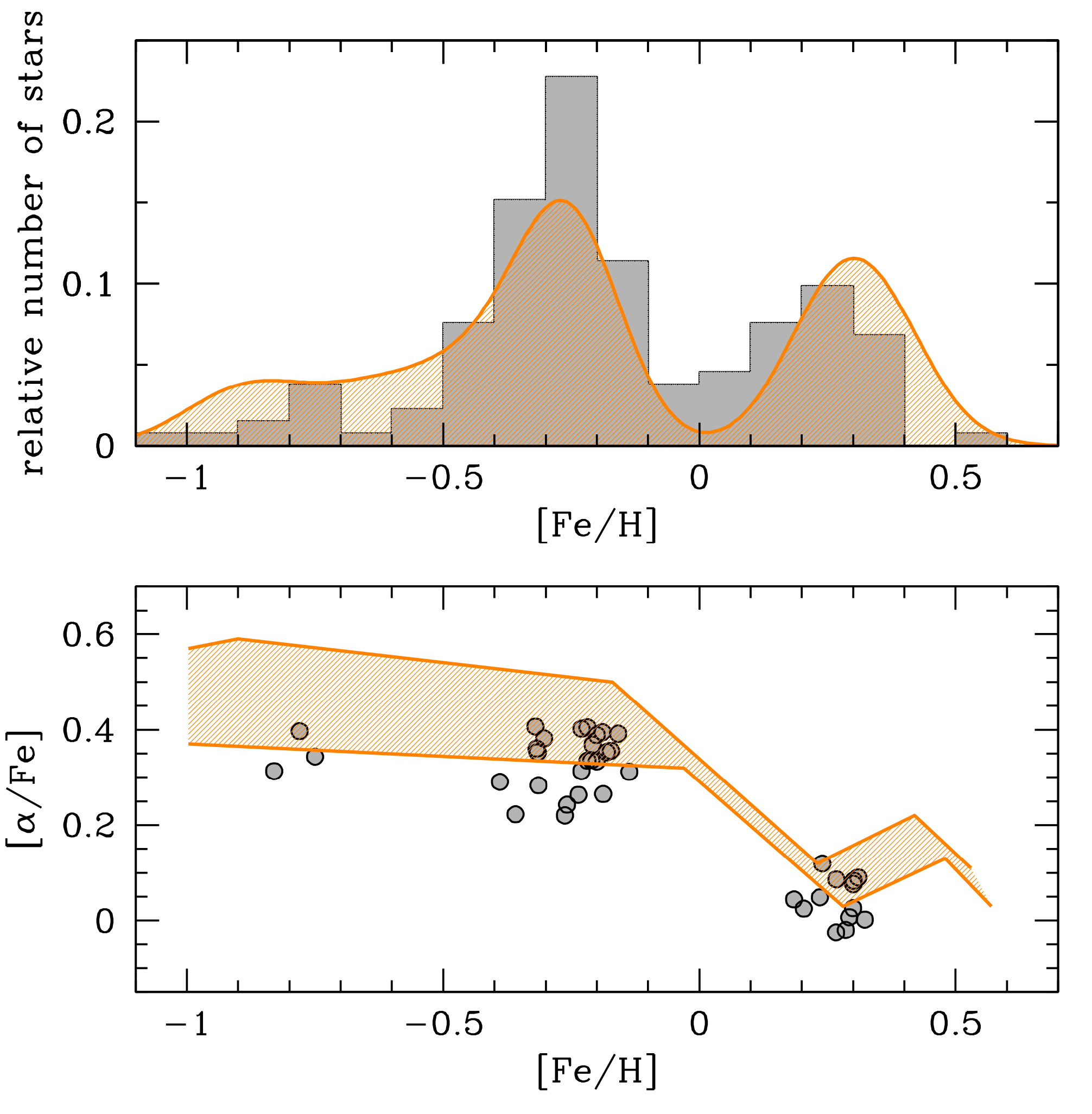}
     \caption{Top panel: Observed iron distribution  of Terzan 5 (gray histogram, from \citealt{massari14}) compared to the prediction of model S02s by \citet[][orange shaded region]{romano23}. Bottom panel: [$\alpha$/Fe]-[Fe/H] abundance pattern observed in Terzan 5 (gray circles, from \citealt{origlia11, origlia13, origlia25}) compared to the prediction of model S02s presented in \citet[][orange shaded region]{romano23}. In both panels, the width of the shaded area represents the uncertainties related to different assumptions about the presence (lower curve) or the absence (upper curve) of hypernovae that, during the explosion, release energies one order of magnitude larger than those of normal core-collapse SNe.}
\label{fig_romano}
 \end{figure}

Among the various models explored by \citet{romano23}, the evolution of a gas clump with an initial mass of $4\times10^7$ M$_\odot$ and a metallicity $Z\sim 0.005$ (Model S02s in their paper) has been found to  reproduce the chemical patterns observed in Terzan 5. As shown in the top panel of Figure \ref{fig_romano}, this model (orange shaded area) quantitatively accounts for the observed iron distribution, which consists in two peaks at sub-solar metallicities (at [Fe/H] $=-0.79$ and [Fe/H] $=-0.3$) comprising about 62\% of the current mass of the system, and a super-solar one (at [Fe/H] $=+0.3$), including the remaining 38\% in mass. Moreover, as shown in the bottom panel of Fig. \ref{fig_romano}, the same model (orange shaded region) correctly reconstructs the enrichment action of SNeII and SNeIa, reproducing the observed abundance pattern in the [$\alpha$/Fe]-[Fe/H] diagram (gray dots, from \citealt{origlia11, origlia13, origlia25}), with the two oldest (12 Gyr old) sub-populations at sub-solar metallicity being enriched only by SNeII (as certified by their $\alpha-$enhancement, [$\alpha$/Fe]$=+0.35$) and the younger (4.5 Gyr old) and super-solar component being enriched by both SNeIa and SNeII (as testified by their solar $\alpha$ content). 

To reproduce the currently observed sub-populations, which comprise $\sim 1.3\times 10^6 {\rm M}_\odot$ of sub-solar metallicity stars and $0.7\times 10^6 {\rm M}_\odot$ of super-solar metallicity stars \citep[see][]{lanzoni10}, the model predicts that the proto-Terzan 5 system formed a total of $5\times10^6$ stars. By assuming a Kroupa (2001) initial mass function (IMF) in the mass range $0.1-100$ M$_{\odot}$, and

\begin{table}
    \scriptsize
    \renewcommand{\arraystretch}{1.25}
    \setlength{\tabcolsep}{9pt}
    \caption{Number of TypeII SNe, neutron stars and black holes expected on the basis of different assumptions for the minimum initial mass to produce a SNII explosion and the minimum initial mass to form a black hole.}
    \begin{tabular}{|c|c|c|c|c|}
    \hline\hline
    minimum mass &  Transition mass  &  N$_{\rm SNII}$  &   N$_{\rm NS}$    &   N$_{\rm BH}$    \\
    SNII explosion     & NS/BH &  &   &   \\
    \hline
    $8 {\rm M}_\odot$  &  $25 {\rm M}_\odot$  & $7.8\times 10^4$ & $6.6\times 10^4$ & $1.3\times 10^4$   \\
    $8 {\rm M}_\odot$  &  $40 {\rm M}_\odot$  & $7.8\times 10^4$ & $7.2\times 10^4$ & $6\times 10^3$   \\
     $11 {\rm M}_\odot$  &  $25 {\rm M}_\odot$  & $4.5\times 10^4$ & $3.8\times 10^4$ & $7\times 10^3$   \\
       $11 {\rm M}_\odot$  &  $40 {\rm M}_\odot$  & $4.5\times 10^4$ & $4.2\times 10^4$ & $3\times 10^3$   \\
    \hline\hline
    \end{tabular}
    \label{tab1}   
\end{table}

that all stars more massive than 8 ${\rm M}_\odot$ explode as SNeII, a total of $7.9\times 10^4$ explosions are predicted. 
Each SNII is expected to leave a compact remnant according to the initial mass of the exploded star: stars less massive than $25 M_{\odot}$ are expected to produce a NS, while more massive stars should generate BHs. In particular (see Fig. 1 in \citealt{heger+03}), stars with initial masses above $40 M_{\odot}$ are expected to directly produce a BH, while stars in the range $25-40 M_{\odot}$ experience too weak explosions to eject most of the material and therefore suffer from 
fallback onto the central NS, which then collapses and forms a BH \citep{macfadyen+01}. 
Under these assumptions, the enrichment process of the proto-Terzan 5 system would produce a total of $6.6\times 10^4$ NSs and $1.3\times 10^4$ BHs.

The vast majority ($\sim 86\%$) of these remnants has been produced by the earliest explosions ($\sim 12$ Gyr ago), while only 14\% of the total is generated during the second main burst, 7.5 Gyr later. Figure 2 schematically shows the contributions of the two main bursts (indicated by the vertical gray strips) to the total number of BHs (top panel) and NSs (bottom panel). 
These numbers can be different if the transition mass is set at $40 M_{\odot}$ (in place of $25 M_{\odot}$), or if the minimum mass for SNII explosion is set to 11 $M_{\odot}$ (instead of $8 M_{\odot}$). Table \ref{tab1} lists the total number of SNeII, NSs and BHs predicted by the model under different assumptions of the minimum mass for SNII explosion and the NS/BH transition mass. The number of NSs varies from about $4\times 10^4$ to $8\times 10^4$, while that of BHs is comprised between $3\times 10^3$ and $13\times 10^3$. In addition, the number of BHs can increase if, as suggested by some simulations \citep{spera+15}, a fraction of stars in the range $18-25 M_{\odot}$ directly form a BH without exploding as SNe. All this clearly introduces some uncertainties in the number of BBH mergers estimated in  Section \ref{sec:BBH}, but no significant impact is expected on the general conclusions concerning the possible significant production of BBH mergers in BFFs.

\section{Estimating the number of binary black hole mergers}
\label{sec:BBH}
The properties outlined in the previous sections suggest that  BFFs are a new class of objects, well distinct form GCs, that could give an exceptional contribution to the number of binary mergers between stellar remnants producing GW emission.
\begin{figure}
    \centering    \includegraphics[width=\columnwidth]{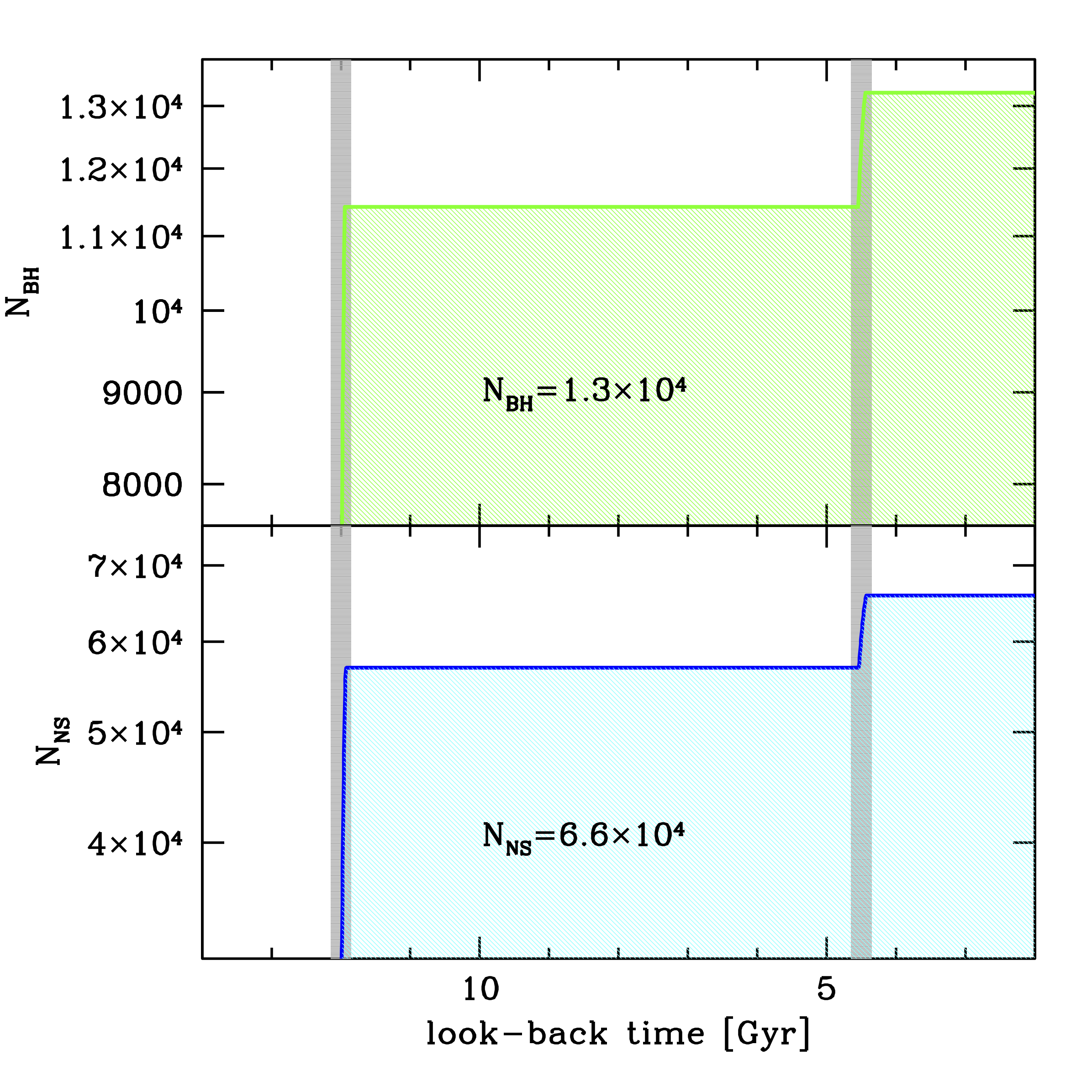}
     \caption{Number of BHs (top panel) and NSs (bottom panel) present within the potential well of Terzan 5 during its evolutionary history, as predicted by model S02s of \citet{romano23}. The two main star formation bursts that characterize the enrichment history of the system (at an age of $\sim 12$ and $4.5$ Gyr) are marked with two vertical gray shaded  strips.}
\label{number}
\end{figure}

\begin{figure*}
    \centering  
   \centering    \includegraphics[width=\textwidth]{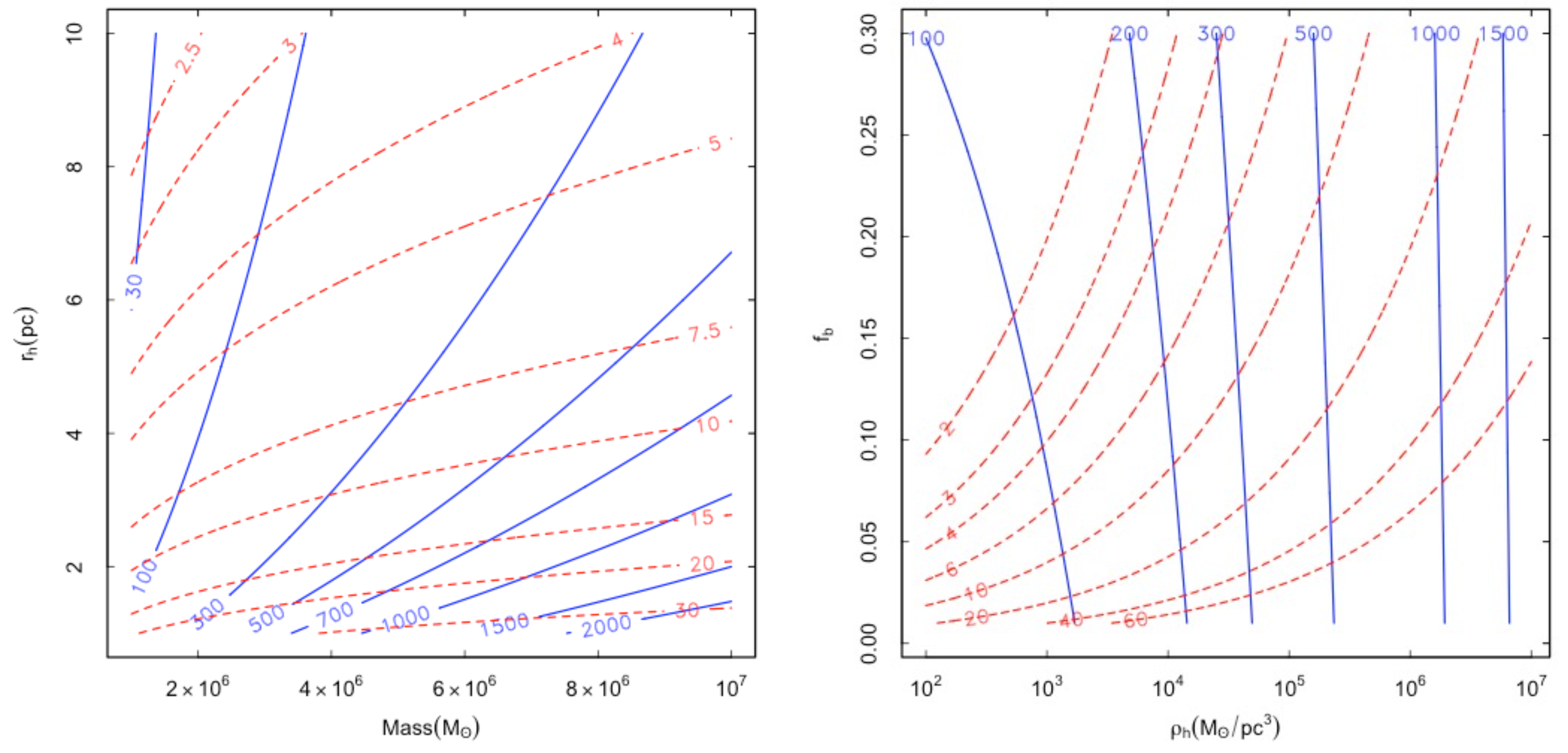}
     \caption{Contour plot of the number of BBH mergers (blue solid lines and associated labels) in the two planes summarizing the effects of the main parameter dependence: initial half-mass radius versus initial stellar mass (by assuming a primordial binary fraction of 10\%; left panel), and primordial binary fraction versus average initial density within $r_h$ (by assuming an initial stellar mass of $3\times10^6  {\rm M}_\odot$; right panel). The dashed red lines and the corresponding red labels show the contour plot of the ratio between the number of dynamical to primordial BBH mergers.}
     \label{mergers}
 \end{figure*}
 
In fact, the large number of SN explosions required to reach the extremely high metallicity regime (half-solar) of the main population currently observed in the two systems, their initial large mass (which allows the retention of dark remnants within the deep potential well after SN explosions), and their high collision rate \citep{lanzoni10, ferraro21}, all concur to favor dynamical interactions and stellar mergers. This suggests that BFFs should be much more efficient GW factories compared to non-collisional stellar structures with similar stellar mass such as dwarf galaxies, which also experienced a smaller number of SN explosions (as testified by their lower star formation rate and lower metallicity). However, recent studies (e.g., Ye et al. 2020 and reference therein) indicate that  dynamics is a very subdominant channel for the production of mergers including NSs. Thus, the following discussion is focused on BBH mergers.

Detailed quantitative estimates of the BFF contribution to the population of BBH mergers would require the modeling of the formation and dynamical evolution of these systems through tailored hydro/N-body simulations able to capture the interplay between the effects of stellar dynamics and multiple star formation episodes, the differences between the  properties of various stellar populations, and the combined effect of such a complex dynamical environment on the system's binary population. While detailed simulations with initial conditions consistent with those expected for the proto-Terzan 5 system and including its major star formation bursts (see Fig. \ref{number}; \citealt{crociati24} and \citealt{romano23}) are currently under construction (Vesperini et al., in preparation), here we provide a first crude estimate of the number of BBH mergers emitting GWs that we expect in Terzan 5 on the basis of its chemical evolution model \citep{romano23} and an extrapolation at high mass of the analytical expressions derived for GCs by \citet{hong18}.

Although this calculation is clearly  an approximation that, as pointed out above, will need to be refined and further investigated with specific numerical simulations, the results obtained in this work provide a first estimate of the contribution to GW emission from the new class of objects represented by the BFFs.

The \citet{hong18} relations have been derived from the analysis of a large survey of Monte Carlo simulations run with the MOCCA code  \citep{giersz13,hypki13} and following the dynamical evolution of GCs for a broad range of different initial properties, such as different initial masses, primordial binary fractions, and half-mass radii (see \citealt{hong18} for further details). The authors calculated the number of BBH mergers produced in each model

and determined how it depends on the clusters' initial conditions. We refer to \citet{hong18} for details on the simulations, but we repeat here the analytical formula they obtained and we use in this paper to estimate of the number of BBH mergers.
From their GC simulation survey, \citet{hong18} found that the total number of BBH mergers ($N_{\rm BBH-merg}$) is well fit by the following analytical expression:
\begin{equation}
N_{\rm BBH-merg}= A \frac{M_0}{10^5 {\rm M}_\odot} \times \left(\frac{\rho_h}{10^5 {\rm M}_\odot {\rm pc}^{-3}}\right)^{\alpha}+B \frac{M_0}{10^5 {\rm M}_\odot} \times f_b, 
\label{eq}
\end{equation}
where $M_0$, $\rho_h$, and $f_b$ are, respectively, the cluster's initial mass, initial mean density within the half-mass radius $r_h$, computed as $\rho_h=0.5 M_0/(4/3 \,\pi r_h^3)$, and primordial binary fraction. The parameters $A$, $B$, $\alpha$ are obtained by fitting the results of the Monte Carlo simulation survey: $A=12.53\pm 0.22$, $B=6.89\pm 0.84$, $\alpha=0.33\pm0.02$. The two terms in the RHS of eq. (\ref{eq}) represent the contribution to the total number of BBH mergers produced by dynamical interactions in the cluster (the first term, proportional to the cluster mass and density within the half-mass radius),  and that due to primordial binaries for which dynamics played no role (the second term, proportional to the cluster mass and primordial binary fraction). 

As discussed in Section 2, the model by \citet{romano23} predicts that the proto-Terzan 5 system had about $5 \times 10^6$ stars, corresponding to a total mass of about $3\times 10^6 {\rm M}_\odot$. Adopting this value for the initial mass, we can use eq. (\ref{eq}) to 
estimate the total number of BBH mergers originating from a system like Terzan 5 for different values of the initial half-mass radius and primordial binary fraction.

If we adopt the relationship between half-mass radius and mass for young massive clusters from \citet{larsen04}, $r_h/\hbox{pc}=2.8  (M/10^4 M_{\rm \odot})^{0.1}$, we find that a proto-Terzan 5 system had an initial half-mass radius of about 5 pc. With these values of the mass and half-mass radius, and assuming a primordial binary fraction equal to 10\%, we find a total of about 135 BBH mergers originating from this system. Adopting the half-mass radius-mass relationship from \citet{marks+12},  $r_h/\hbox{pc}=0.33 (M/10^4 M_{\rm \odot})^{0.13}$, we find a significantly more compact system, with just $r_{\rm h} \sim 0.7$ pc, which would produce an even larger number of BBH mergers, equal to about 835. 
 
To take into account that systems like the progenitor of Terzan 5 might follow different half-mass radius-mass relationships, and to provide a more general illustration of the expected number of BBH mergers for a broader range of possible initial properties, we show in Fig. \ref{mergers} the contour plots of the number of BBH mergers (blue solid lines) in the plane of initial half-mass radius versus mass, assuming a 10\% primordial binary fraction (left panel), and in the plane of primordial binary fraction versus $\rho_h$, assuming $M_0= 3 \times 10^6 {\rm M}_\odot$ (right panel). 

The two panels of this figure show the dependence of the  number of BBH mergers on the binary fraction, mass and half-mass radius (and thus half-mass density). As illustrated by this figure, the number of BBH mergers is particularly  sensitive to the adopted $r_h$-mass relation, which also sets the central density. This is the reason why we assumed two different expressions for this relation (from \citealt{larsen04} and \citealt{marks+12}) likely delimiting a realistic range of half-mass radii for a proto-Terzan  5 object. Given the uncertainty in the initial structural parameters of the system, the contour values
shown in the figure should be considered as a plausible order-of-magnitude estimate, rather than a precise prediction, suggesting a total number of BBH mergers that approximately varies  between $10^2$ and $10^3$.

As discussed above, the analytical expression obtained by \citet{hong18} includes a contribution from dynamical BBH mergers and one from primordial BBH mergers. Fig. \ref{mergers} 
shows the ratio of the number of dynamical to primordial BBH mergers (red dashed lines and labels) calculated as the ratio of the two terms in eq. (\ref{eq}) extrapolated to the case of a BFF. For proto-Terzan 5 systems, our calculations 
indicate that the dynamical channel of BBH merger formation dominates significantly over the primordial one, with values of the dynamical to primordial BBH merger number ratio ranging from $\sim 3$ for the low-mass low-density systems, to $\sim 30$ or even more for the most massive and high-density systems. This result further illustrates the key role played by internal dynamical evolution, and its interplay with the evolution of the binary population in the production of BBH mergers.

\begin{figure}[h]
    \centering    \includegraphics[width=\columnwidth]{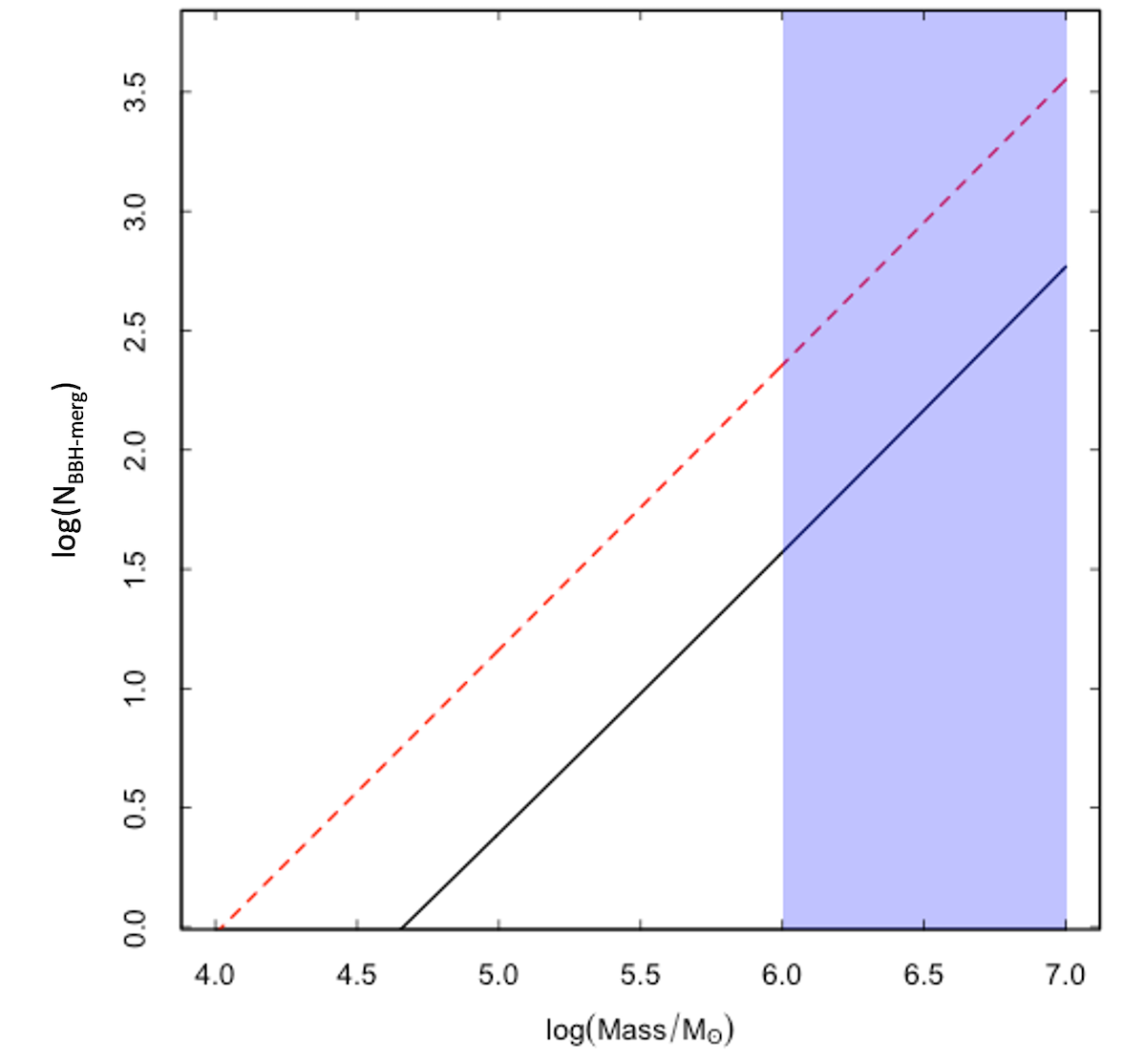}
     \caption{Number of BBH mergers as a function of the system initial stellar mass from eq. (\ref{eq}),  assuming the half-mass radius-mass relationship from \citet[][solid line]{larsen04} and that from \citet[red dashed line]{marks+12}. The purple shaded area marks the range of possible initial masses for proto-Terzan 5 systems. 
     }
     \label{mergers2}
 \end{figure}

\begin{figure}[h]
    \centering    \includegraphics[width=\columnwidth]{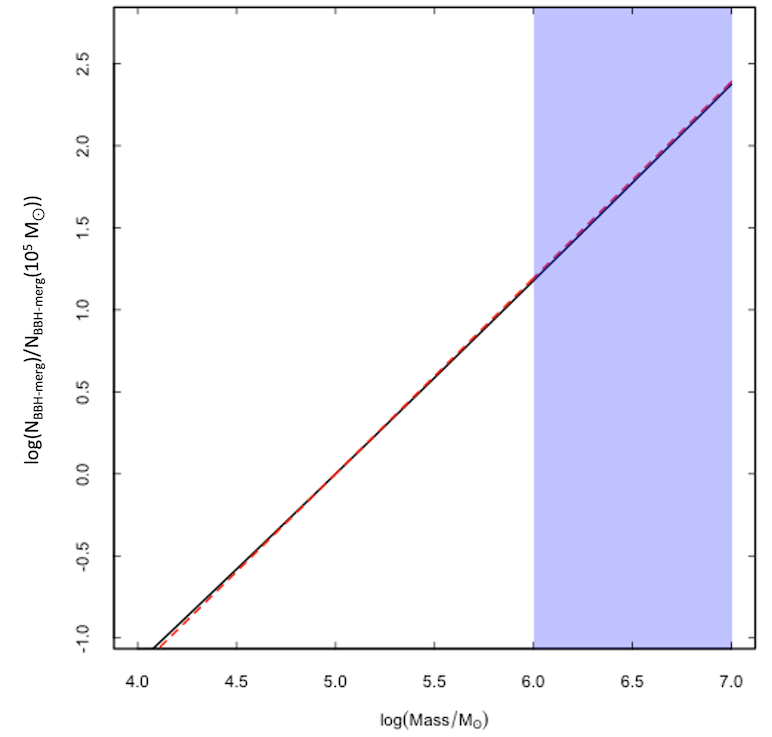}
     \caption{As in Figure \ref{mergers2}, but for the number of BBH mergers normalized to that expected in a system with mass equal to $10^5 {\rm M}_\odot$.}
     \label{mergers3}
 \end{figure}
 
\section{Discussion and conclusions}
The chemical evolutionary history together with the reconstructed star formation history of Terzan 5 \citep[][respectively]{romano23, crociati24} suggest that this system has formed a very large population of degenerate objects (see Fig. \ref{number}) as the result of the large number of SNII explosions required to enrich gas in iron while maintaining the [$\alpha$/Fe] ratio high, consistently with the spectroscopic observations. By using the total number of stars formed in this stellar system during its enrichment process (from \citealt{romano23}), and the relations computed by \citet{hong18} for a large sample of simulated GCs with different initial conditions, we provided a first quantitative estimate of the number of BBH mergers in Terzan 5. As illustrated in Fig. \ref{mergers}, we find that several hundreds of BBH mergers are expected, 
especially in conditions of high initial density and mass, and from the dynamical channel.
Indeed, as shown in Fig. \ref{mergers2}, the  number of BBH mergers obtained for the mass range representative of proto-Terzan 5 systems (purple strip in the figure) is significantly larger than that expected in ``normal'' GCs, which typically have masses below $10^6 {\rm M}_\odot$. In Fig. \ref{mergers3} we show the number of BBH mergers normalized to 
that expected in a GC with a mass of $10^5 {\rm M}_\odot$. It is interesting to note that, assuming mass values ranging from $10^6 {\rm M}_\odot$ to $10^7 {\rm M}_\odot$ as possible initial masses for proto-Terzan 5 systems, the number of BBH mergers produced in these objects is between $\sim 15$ and $\sim 250$ times larger than that produced by a GC with a mass of $10^5~{\rm M}_\odot$. This implies that a BFF \emph{alone} with an initial mass of $\sim 10^7 {\rm M}_\odot$ is expected to contribute a number of GW sources larger than that expected from the \textit{entire} present-day population of all Galactic GCs (which are approximately 150 in the Milky Way; \citealt{harris}). 

This demonstrates the relevance  of this new class of stellar systems in the production of GWs. 

In spite of the approximations used in the presented calculations,
this study suggests that the population of BFFs is a new and very promising Galactic factory of GW sources. This implies that the building blocks of galaxy bulges (the BFFs) could play a dominant role in the production of GWs along all cosmic epochs. In fact, under the hypothesis that galaxy bulges form through the merger of primordial massive systems \citep[e.g.,][]{imm04, elme08, bournaud09}, we expect that hundreds of proto-BFFs were present at the epoch of the Galaxy assembling in order to provide the entire mass budget of the Milky Way bulge ($\sim 2\times10^{10} {\rm M}_\odot$; \citealt{valenti16}). 
Thus, this investigation unveils a new additional population of GW emitters that was not considered in any previous study and shows that the building blocks of galaxy bulges are factories of a significant number of sources of GW emission in spirals.

An additional consideration that can be derived from this study concerns the formation of massive stellar BHs (with masses larger than 50-60 ${\rm M}_\odot$, as those detected with LIGO/VIRGO observations; \citealp{abbott+23b}) and even intermediate-mass BHs (IMBHs). In fact, 
the deep potential well of the BFF progenitors could have retained a large fraction of degenerate stellar remnants, much larger than for genuine GCs. This is indeed testified by the huge number of X-ray sources and millisecond pulsars detected in Terzan 5 (\citealt{heinke06, ransom05, cadelano18, pad24}), which could be only the tip-of-iceberg of a larger population of objects that are not yet detected. Moreover, the fact that these degenerate remnants are more massive than normal cluster stars guarantees that they rapidly migrate toward the center. On the other hand, the high collisional rate of the system (Terzan 5 shows the largest value of the entire Galactic GC population; see \citealt{verbunt87, lanzoni10}) guarantees a large number of binary-binary interactions, thus favoring the formation of close and heavily degenerate binaries. This scenario would suggest that a BFF like Terzan 5 could provide the optimal conditions for recurrent mergers of BBHs with increasing masses, which could lead to the formation of an IMBH \citep[e.g.,][]{giersz15, dicarlo21}. Once applied to primordial proto-galaxies, this mechanism may provide new insights into BH growth in the early Universe.

Hence, as a general result, this study identifies BFFs not only as very efficient sources of GWs, but also as the natural place where massive stellar BHs and IMBHs can form via repeated dynamical interactions, thus further increasing the importance of this class of stellar systems in the context of the modern astrophysical research. 

Due to the approximations adopted in this study, the presented results should be considered just as a first general estimate of the possible contribution of the BFF family to the population of BBH mergers. The estimate has been obtained by using a relation determined from a survey of GC simulations \citep{hong18}, which predicts the number of BBH mergers as a function of the total mass, average half-mass density, and binary fraction of the host stellar system (see eq. 1). Under the assumption that this relation holds in the mass range of the proto-Terzan5 system, we used the total mass (in stars) predicted by the chemical model of \citet{romano23}, thus implicitly taking into account the entire enrichment history of the system. 
While the binary fraction has a negligible effect (see the right panel of Fig. \ref{mergers}), the major limitations come from the uncertainties in the parameters characterizing the proto-Terzan 5 system, which have large impact on the final estimates (see Figs. \ref{mergers} and \ref{mergers2}), and possible differences in terms of internal dynamical evolution between a genuine GC and a BFF. Hence, while some differences may be expected, these cannot be predicted a-priori and new simulations tailored to model more in detail the formation and dynamics of BFFs are needed. In future investigations we will build specific models for BFFs and consider how the complex star formation history and different metallicities of the stellar populations of these systems, along with other dynamical processes (e.g., the formation of an IMBH; see \citealp{hong20}) affect the number of BBH mergers produced by BFFs (Vesperini et al., in preparation). Moreover, we are
also planning to expand the investigation of the origin of these
systems in the context of cosmological simulations, to infer their
star formation history and the distribution of their initial properties
(Calura et al., in preparation).

 \begin{acknowledgements}
This work is part of the project  GENESIS 
"Searching for the primordial structures of the Universe in the heart of the Galaxy" (Advanced Grant FIS-2024-02056, PI:Ferraro), funded by the Italian MUR through the Fondo Italiano per la Scienza (FIS) call. ED acknowledges financial support from the INAF Data Analysis Research Grant (Ref: Dalessandro) of the ``Bando di Astrofisica Fondamentale 2024''.
\end{acknowledgements}

\bibliographystyle{aa}

\vspace{5mm}

\end{document}